\documentclass{sig-alternate}

\permission{\copyright 2017 International World Wide Web Conference Committee \\ (IW3C2), published under Creative Commons CC BY 4.0 License.}
\conferenceinfo{WWW'17 Companion,}{April 3--7, 2017, Perth, Australia.}
\copyrightetc{ACM \the\acmcopyr}
\crdata{978-1-4503-4914-7/17/04. \\
http://dx.doi.org/10.1145/***Add your assigned DOI** \\
\includegraphics{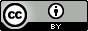}
}

\newdef{definition}{\textbf{Definition}}
\newdef{assumption}{\textbf{Assumption}}

\usepackage{tikz}
\usepackage{amssymb,bbm,amsmath,url,mathtools}
\usepackage{siunitx,subfig,multirow,amsmath,pgfplots,adjustbox,wasysym,marvosym}
\usepgfplotslibrary{groupplots}
\usetikzlibrary{patterns}
\usepackage[ruled,linesnumbered]{algorithm2e}

\newcommand{\specialcell}[2][c]{%
  \begin{tabular}[#1]{@{}c@{}}#2\end{tabular}}

\def\addlegendimage{\csname pgfplots@addlegendimage\endcsname}
\DeclareMathOperator*{\argmin}{arg\,min}
\DeclareMathOperator*{\argmax}{arg\,max}
\SetKwInOut{Parameter}{Parameter}

\begin{document}

\title{Numerical Facet Range Partition: Evaluation Metric and Methods}

%
%
%
%
%

\numberofauthors{2} 
%
\author{
\alignauthor
Xueqing Liu, Chengxiang Zhai\\
       \affaddr{University of Illinois at Urbana-Champaign}\\
       \affaddr{Urbana, IL 61801}\\
       \email{xliu93,czhai@illinois.edu}
\alignauthor
Wei Han, Onur Gungor\\
       \affaddr{WalmartLabs}\\
       \affaddr{Sunnyvale, CA 94806}\\
       \email{whan,ogungor@walmartlabs.com}
}

\maketitle
\begin{abstract}
Faceted navigation is a very useful component in today's search engines. It is especially useful when user has an exploratory information need or prefer certain attribute values than others. Existing work has tried to optimize faceted systems in many aspects, but little work has been done on optimizing numerical facet ranges (e.g., price ranges of product). In this paper, we introduce for the first time the research problem on numerical facet range partition and formally frame it as an optimization problem. To enable quantitative evaluation of a partition algorithm, we propose an evaluation metric to be applied to search engine logs. We further propose two range partition algorithms that computationally optimize the defined metric. Experimental results on a two-month search log from a major e-Commerce engine show that our proposed method can significantly outperform baseline.
\end{abstract}

\keywords{Faceted search; User search log; Information retrieval models; Non-smooth optimization}


\section{Introduction}
\label{sec:intro}

\begin{figure}
\centering
\includegraphics[scale=0.25]{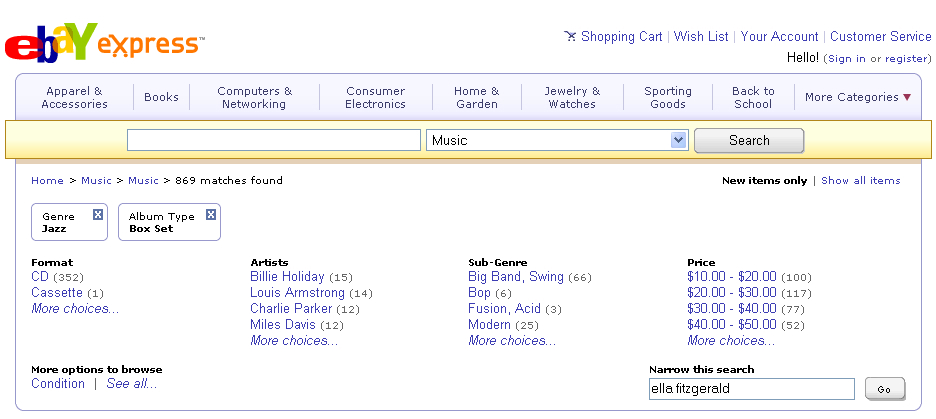}
\caption[Caption for LOF]{Snapshot of faceted search system on eBay, picture borrowed from Hearst~\cite{hearst2009search} (Figure 8.12, page 195)\label{fig:intro_example}}
\vspace{-0.2in}
\end{figure}

Querying and browsing are two complementary ways of information access on internet. As one convenient tool to help browsing, faceted search systems have become an indispensible part of today's search engines. Figure~\ref{fig:intro_example} shows a standard faceted system on eBay. Upon receiving user query, it displays a ranked list of \emph{facet}s: format, artists, sub-genre and price, along with facet values under each facet. These facet values are metadata of the search results. When user selects one or more values, search results are refined by the selection, e.g., in Figure~\ref{fig:intro_example}, the results (not displayed) only contain box set albums whose genres are Jazz. Faceted browsing is largely popular in search engines for structured entities of the same type\footnote{In this paper, we frequently use the term `entity' to refer to any structured entity. We do not use the term `item' because the search object we study is more general, e.g., people search. In the experiment part where our data is from e-Commerce engine, we use the term `product' instead.} (e.g., e-Commerce products, movies, restaurants). In these engines, user often lacks the ability to specify facet values in detail \cite{1555452}. Therefore, faceted system such as Figure~\ref{fig:intro_example} can serve as a convenient tool to elicit user's needs so they can quickly click on the suggested facet values to expand their queries. Faceted browsing is also exceedingly helpful on touch screen devices, where typing query is less convenient than clicking on a facet.

A faceted system consists of multiple components, which would naturally decompose its optimization into multiple sub-problems. Existing works have covered quite a few of these sub-problems, e.g., ranking facets or values~\cite{Zwol,journals/ijon/KangYZTHC15,kashyap10}, facet selection~\cite{Liberman:2012:AOF:2245276.2245409,Roy08}. However, we identify one problem which, to the best of our knowledge, has never been formally studied before. Basically, how to suggest values of a \emph{numerical facet} to help user browse the query results? An example of numerical facet is price in Figure~\ref{fig:intro_example}, where the result albums are partitioned into 5 non-overlapping subsets based on their prices: [0, 20), [20, 30), [30, 40), [40, 50) and [50, $\infty$). This is equal to saying the results are separated by 20, 30, 40 and 50. So the problem is rephrased as: given user query and results, how to find the best separating values? This problem has a clearly different goal from existing works in faceted system~\cite{Roy08,Liberman:2012:AOF:2245276.2245409,kashyap10,Koren:2008:PIF:1367497.1367562,Hearst2008UI,Vandic:2013:FSA:2505515.2505664}. It can be further decomposed into two parts. First, how to evaluate the quality of a set of separating values (e.g., how good is 20,30,40 and 50?)? Second, if we can find such a metric, how to find separators that optimize it? 

\begin{figure}
\centering
\includegraphics[scale=0.5]{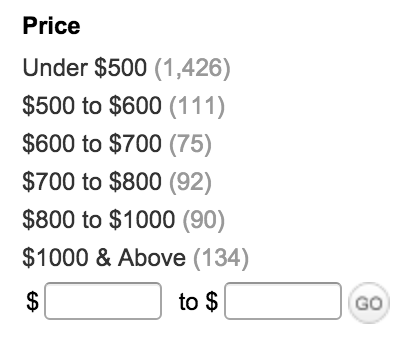}
\caption{A specific example of the `one range dominates' issue (Table~\ref{tab:websites}). The snapshot was taken on 01/21/2016, on Amazon under query `refurbished laptop'.\label{fig:intro_example_3}}
\vspace{-0.2in}
\end{figure}

Before we delve into answering the two questions, one may wonder why it is even important to study this problem. Arguably, numerical facets are only a small portion of all facets, and why are we unhappy with the current design? If we only consider one search engine, indeed, it usually just contains one or a few numerical facets (e.g., Figure~\ref{fig:intro_example}). However, notice numerical facets span a wide range of applications. Some of the examples are news search (timestamp), location search (distance), e-Commerce search (price, mileage, rating) and academic search (h-index). So focusing on numerical facet does not make our study narrow. For the latter question, we conduct a case study on the price ranges from top-10 shopping websites that provide price suggestion\footnote{Ranking is based on the website traffic statistics from \url{www.alexa.com} as of 02/16/2017.}. We find several issues which we demonstrate in Table~\ref{tab:websites} and Figure~\ref{fig:intro_example_3}. The most common issue is that among multiple suggested ranges, one range contains the majority of results, e.g., Figure~\ref{fig:intro_example_3} shows that range [0, 500) contains 73.9\% of the products under query `refurbished laptop'. It can be expected that the majority users would click on [0, 500), but this only reduces the total number from 1,928 to 1,426, which does not seem very helpful. Another issue we find on one website (\url{www.etsy.com}) is it appears to suggest fixed ranges (25, 50, 100) for all queries, so it is not adaptable to different queries such as `dress' and `hair pins'. Finally, the price ranges from eBay appear to be the most adaptable among all 10 websites, but seems its number of ranges is fixed to 3, making it unable to adapt to price-diversified categories such as camera. Based on the study results, we believe there is still plenty of room for improving range partition techniques in current search engines. 
\begin{table}
\centering
\begin{tabular}{|l|c|c|}
\hline
website & issue & example query\\ \hline
\url{amazon.com} & one range dom. & refurbished laptop\\
\url{ebay.com} & 3 ranges & laptop; camera\\ 
\url{walmart.com} & one range dom. & socks\\ 
\url{bestbuy.com} & one range dom. & phone charger\\ 
\url{etsy.com} & fixed ranges & dress; hair pins\\ 
\url{homedepot.com} & one range dom. & french door fridge\\ 
\url{target.com} & one range dom. & card game\\ 
\url{macys.com} & one range dom. & soap\\ 
\url{lowes.com} & one range dom. & pillow\\ 
\url{kohls.com} & one range dom. & socks\\ \hline
\end{tabular}
\caption{Issues of suggested price ranges among top-10 shopping websites (as of 02/16/2017).\label{tab:websites}}
\end{table}

For the first question, we evaluate our problem by collecting past user search log and defining our evaluation metric on top of it. It is a common practice in information science to evaluate an information system using user's gain and cost~\cite{card:1999,conf/sigir/Azzopardi14,conf/cikm/YilmazVCRB14}, where the gain is often estimated as the (discounted) number of relevant entities or clicks in the log~\cite{journals/tois/MoffatZ08}, and cost is often estimated as the total number of viewed entities in the log~\cite{Liberman:2012:AOF:2245276.2245409}. Similarly, evaluation metric for a set of numerical ranges can be defined as user's cost and gain when using the ranges to browse the results. Following existing works in faceted system~\cite{Liberman:2012:AOF:2245276.2245409}, we fix the gain to 1 and use the cost as our evaluation metric. Under a few reasonable assumptions (Section~\ref{sec:reusing}), the cost is equal to the rank of the first clicked entity (in the log) in the unique range (among the set of ranges) that contains it.

After the first question is answered, we shift our focus to the optimization problem. From examples in Figure~\ref{fig:intro_example} and Figure~\ref{fig:intro_example_3}, we can observe that a good partition should (at least) satisfy the following properties: first, it is good for the suggested separators to be adaptable to each query; second, instead of letting one range dominates, the number of entities in each range should be more balanced; third, our partition algorithm should be able to generate any number of ranges, instead of only one specific number like 3. There exists a simple solution that satisfies all three properties: just partition the results into $k$ ranges, so that each range contains the same number of entities. We call this simple method the \emph{quantile} method. Indeed, the quantile method reduces the maximum cost in Figure~\ref{fig:intro_example_3} from 1,426 to 321. But can we further improve it?

In this paper, we propose two range partition algorithms. The idea is to collect a second search log and use it for training, to help improve the performance on the search log for evaluation. In the first proposed method, training data is used for estimating the expected click probabilities in the testing data, then the range is computed by optimizing the expected cost using dynamic programming; in the second method, we propose to parameterize the problem and optimize the parameters on the training data. We conduct experiments on a two-month search log collected from Walmart search engine. Results show that our method can significantly outperform the quantile method, which verifies that learning is indeed helpful in the range partition problem.

\def\ignore#1{}
\ignore{Suppose otherwise we partition the two examples into equal sized ranges, each range will then contain 321 and 408 entities. In that case, the user can avoid having to go through 1,500 entities. The even partition is also named \emph{equi-depth bining} or \emph{quantile} method in literature~\cite{conf/sigmod/MuralikrishnaD88}. \\
So we know quantile method is better than Figure~\ref{fig:introeg} in terms of browsing cost. But how exactly do we quantify the browsing cost of a partition? Previous literatures have provided some hints, where the browsing cost is defined as the number of entities above the clicked one \cite{Liberman:2012:AOF:2245276.2245409}, i.e., the rank of clicked entity. However, no previous work has ever defined or evaluated the browsing cost for the numerical partition. \\
The evaluation of a faceted search system is in general a challenging task~\cite{twitter}. This is because today's search engines allow users to select multiple facets at the same time, so the browsing cost log may not just reflect the performance of a single facet, and cannot be easily decomposed. Although we can do online testing such as A/B testing, it is expensive and this problem still exists. To make this setting more accessible and to make the problem more general, we still try to evaluate the browsing cost in an offline manner. We will leave the online testing problem for future work.\\
In an offline evaluation, we have to make assumptions on the facet user chooses. Previous work~\cite{Liberman:2012:AOF:2245276.2245409} on facets selection has made assumption that user is always able to select the facet which minimizes her browsing cost. Although this is a strong assumption on user, which is not necessarily true in real world scenario, it also happens to be one of the only few possible ways for offline evaluation. We follow \cite{Liberman:2012:AOF:2245276.2245409} and assume user would always be able to first click on the range(s) that contain the relevant entity(s). Notice this is actually not a very strong assumption when the user knows more or less about her price need. Meanwhile, it is almost impossible to predict user's abandonment on a facet range. This assumption is thus the most effective way we can think of to fully leverage search log and fairly compare different methods.\\
Based on this assumption, we study multiple methods for generating the partition: first, we study a dynamic programming approach to directly optimize the expected browsing cost during testing. Dynamic programming has been applied in related work on database systems~\cite{conf/vldb/JagadishKMPSS98}, but not on numerical partitioning; second, we propose a machine learning based method by optimizing the partition parameter on training data. We empirically compare the two methods and the aforementioned quantile method }


\pdfoutput=1

\section{Related Work}
\label{sec:relwork}

During the past decades, researchers design different interfaces for faceted search and browsing. They include faceted system that displays one facet~\cite{Roy08} and $k$ facets~\cite{Liberman:2012:AOF:2245276.2245409,Vandic:2013:FSA:2505515.2505664}, where the facet selection is based on ranking. Due to the heterogeneity of entity structures on the web, facets ranking can be classified as ranking facet~\cite{BasuRoy:2008:MDD:1458082.1458088}, ranking facet values~\cite{journals/ijon/KangYZTHC15} and ranking (facet, value) pairs~\cite{kashyap10}. There are also faceted systems which support image search~\cite{Zwol} and personalized search~\cite{Koren:2008:PIF:1367497.1367562}. To the best of our knowledge, we have not found any existing literature that explains how to suggest numerical ranges that are adaptable to user queries. 

It is a common practice to evaluate search engine using user's gains and costs~\cite{järvelin2002cumulated,journals/tois/MoffatZ08,conf/sigir/Azzopardi14,conf/cikm/YilmazVCRB14}. Existing approaches would define a system's utility as the difference between user's gain and cost~\cite{card:1999,journals/tois/MoffatZ08}, or they would evaluate gain and cost separately~\cite{conf/sigir/Azzopardi14,conf/cikm/YilmazVCRB14}. Meanwhile, existing works in faceted systems have also defined metrics for self evaluation~\cite{Liberman:2012:AOF:2245276.2245409,kashyap10,BasuRoy:2008:MDD:1458082.1458088}. \cite{Liberman:2012:AOF:2245276.2245409} defines the metric as rank of the relevant document after user selects some facets; \cite{kashyap10} instead defines it as the total number entities after user selects facets. Between the two, we believe the former one better reflects the actual user cost, so we choose to use it in our metric (Section~\ref{sec:metric}), although the latter one is easier to compute. 

Since faceted system is an interactive environment, it is usually impossible to collect the actual user behavior on the system to test. As a result, almost all the evaluation in faceted system have to rely on making assumptions to approximate user behavior~\cite{Roy08,Liberman:2012:AOF:2245276.2245409,kashyap10}. For example, \cite{Liberman:2012:AOF:2245276.2245409} tests two assumptions: (1) user would (conjunctively) select all facets that helps to reduce the rank of relevant document; (2) user would only select the facet that reduces the most of this value. \cite{Roy08} assumes the user would follow the behavior they estimated from 20 users in a pilot study on a different environment. \cite{conf/sigir/ZhangZ15} assumes the probability for user to select each facet is proportional to the semantic similarity between the facet and the relevant document. Unlike \cite{conf/sigir/ZhangZ15}, our assumption in Section~\ref{sec:reusing} only relies on user's discminative knowledge on facet values, and unlike \cite{Liberman:2012:AOF:2245276.2245409}, we do not make further assumptions on user's knowledge about data distribution. So our work relaxes the assumptions made by previous works. 

Our problem is remotely related to generating histograms for database query optimization~\cite{conf/vldb/JagadishKMPSS98,conf/pods/AcharyaDHLS15,conf/sigmod/MuralikrishnaD88}. Different from our query adaptive ranges, histograms are used for data compression so they are fixed for all queries. Same as our first method (Section~\ref{sec:firstmethod}), Jagadish et al.~\cite{conf/vldb/JagadishKMPSS98} also uses dynamic programming, although for a different optimization goal. Recently, \cite{conf/pods/AcharyaDHLS15} leverages an approximation technique and is able to replace DP with a linear time algorithm. However, this approximation technique is not applicable to our case, simply because we have a different optimization goal. Our first method would remain a super-cubic running time.

\vspace{-0.13in}

\pdfoutput=1

\section{Formal Definition}
\label{sec:problem}

We formally define the numerical range partition problem and introduce notations that we will use throughout the rest of the paper. Suppose we have a working set of entities $E=\{e_1, \cdots, e_{|E|}\}$ that user would like to query on. Each entity $e\in E$ is structured, meaning it contains one or multiple facets. For example, facet values of one specific laptop entity is: Brand=Lenovo, GPU=Nvidia Kepler, etc. Here `Brand' and `GPU' are facets; `500GB' and `Nvidia Kepler' are facet values. Facets are often shared by entities in $E$, but some facets are only shared by a subset of $E$. For example, some laptops do not have a GPU. 


At time $i$ user enters a query $q^i$, search engine retrieves a ranked list of entities $E^i\subset E$. Our problem asks, for one specific numerical facet (e.g., price), how to find a set of separating values for that facet? In order for this problem to exist, at least a significant number of entities in $E^i$ should contain the specified numerical facet. From now on, we will just assume this facet is already specified and all the discussions are about this facet. 

We further assume the number of output ranges is given as an input parameter $k$. $k$ is defined by either system or user. We believe it is important to have control on the number of output ranges. Indeed, it would be bad experience if the user wants to see fewer ranges but receives an unexpectedly long list. Also, it is unfair to compare two partition algorithms if they generate different number of ranges, e.g., [0, 100), [100, 200), [200, 300), [300, 400) is almost certainly better than [0, 200), [200, 400) because user can always use the former one to zoom into a better refined results. 

To summarize the input and output of a range partition algorithm: \textbf{Input}: (1) number of output ranges $k$; (2) query $q^i$; (3) ranking algorithm and ranked list $E^i$; numerical facet value of each $e\in E^i$, denoted as $v(e)$ (if $e$ does not have the facet, $v(e)$ is empty); rank of each $e\in E^i$, denoted as $rank(e)$. \textbf{Output}: $k-1$ separating values $S^i = (s_1, \cdots, s_{k-1})\in \mathbb{R}^{k-1}$, where $s_1 < \cdots < s_{k-1}$.

\section{Evaluation}

In this section, we propose and formally define our evaluation techique and metric for range partition algorithms. 

\subsection{User Behavior Assumptions}
\label{sec:reusing}

Evaluation in IR is mainly divided into two categories: first, conduct user studies such as laboratory based experiments or crowdsourcing; second, collect search log of real user engagements in the past, define evaluation metrics on top of the log and use them to compare different systems' performances, also called Cranfield-style evaluation~\cite{SparckJones:1997:RIR:275537}. Since the former approach is expensive and not easy to reproduce, we choose the latter one, which is also the more frequently used approach of evaluating faceted systems in existing work~\cite{Liberman:2012:AOF:2245276.2245409,Vandic:2013:FSA:2505515.2505664}. Collected log consists of queries, and we only keep queries with at least one clicked entity. Also in this paper, we assume user click is the only relevance judgement. That is, \emph{relevant entity is equal to clicked entity}. 

But it is not straightforward how to obtain a reusable search log for evaluating range partition algorithms. On the one hand, it is impossible for the search log to have enumerated all possible range sets. On the other hand, unlike reusable relevance judgements in Cranfield experiments, it is difficult to infer which range user would select out of one set based on her selection out of a different set in the log. Fortunately, existing work in faceted search~\cite{Liberman:2012:AOF:2245276.2245409} provides a hint to this challenge. It assumes user would be able to select the facet value that is most helpful in reducing the rank of the relevant document, then sequentially browse the refined document list until finding the relevant document. In other words, it assumes \emph{user has some partial knowledge in which facet value is more relevant before actually seeing the relevant document}. Similarly, we can assume:

\begin{itemize}
\vspace{-0.02in}
\setlength\itemsep{0em}
\item \textbf{Assumption 1}. User would select the range that contains the relevant entity;
\item \textbf{Assumption 2}. After selecting the relevant range, user would sequentially browse the refined results until reaching relevant entity;
\vspace{-0.02in}
\end{itemize}

Assumption 1 only requires user has a discriminative knowledge on the numerical facet (e.g., knowing which price range is more relevant); while Assumption 2 is among the basic assumptions of information retrieval~\cite{Craswell:2008:ECC:1341531.1341545,robertson-the-1997}. 

There are cases where our assumptions may not be true. For example, if the numerical value of relevant entity is near the borderline, it is difficult for the user to choose between the two ranges. However, we find them reasonable to make when our main purpose is to perform comparative studies between different partitioning algorithms. This is because if there is any bias introduced through these assumptions, the bias is unlikely favoring any particular algorithm. 


\subsection{Evaluation Metric}
\label{sec:metric}

\newcommand\rrank{\mathop{\mbox{$Refined$-$Rank$}}}
It is a common practice in information science to evaluate a system's performance using user's cost and gain. Previous evaluation methods can be categorized into three groups. First, evaluate cost and gain separately~\cite{conf/cikm/YilmazVCRB14}. Since our goal is comparative study, this approach is not informative enough. Second, use the difference between gain and cost, e.g., gain divided by cost~\cite{card:1999}. Although thereby we only have one score, this approach will likely introduce bias since gain and cost may not be on the same scale. The third approach is to control one variable while examining the other. In our problem, it is easier to control and measure gain, since it can be simply defined as the number of entities user has clicked so far. Meanwhile, reusing search log has added challenge to measuring cost of faceted system. Although cost in a no-facet search engine can be simply estimated as number of entities above relevant ones; in engines with faceted system, however, if the number of relevant entities (i.e., user clicks) is larger than 1, this definition is ambiguous, because there are many possible cases of user activity, and cost in each case is different \footnote{For example, under one query, user clicked on entity $e_a$ and $e_b$, and they are in range $a$ and $b$ (different). Case 1: user selects both $a$ and $b$,  browse until finding both $e_a$ and $e_b$. Case 2: user selects $a$, browse until finding $e_a$, unselect $a$ and select $b$, browse until finding $e_b$. Case 3: user selects $a$, browse until finding $e_a$, select $b$, browse until finding $e_b$. }.

On the other hand, if the number of clicked entities is fixed to 1, i.e., we only consider the first clicked entity in the log, it is easy to obtain an unabmiguous definition for cost: for any suggested ranges, there will be one and only one range that contains the relevant (clicked) entity. So if we apply the two assumptions in Section~\ref{sec:reusing}, user would first select that unique range, then sequentially browse entities in that range until finding the first relevant entity. Therefore, the cost is equal to the rank of the first clicked entity in its unique range. We assume that after user selects any range, relative ranks of entities inside that range do not change. Therefore the cost is well defined by the initial search results list $L$, the suggested range $S\in \mathbb{R}^{k-1}$ and the first clicked entity $e$, we denote this value as $\rrank(e, L, S)$.

Now we are ready to define the evaluation metric for a range partition algorithm $A$. At time $i$ in the log, user enters query $q^i$, search engine returns ranked list $E^i$ and user first clicked on entity $e^i$. Suppose algorithm $A$ suggests ranges $S^i=(s_1, \cdots, s_{k-1})$ for each query $q^i$ in the log, we evaluate algorithm $A$'s performance using the \emph{averaged refined rank} metric, or ARR for short:
\vspace{-0.02in}
\begin{eqnarray}
RR_i &=& \rrank(e^i, E^i, S^i)\nonumber\\
ARR &=& \frac{1}{n}\sum_{i=1}^n RR_i\label{eq:arr}
\end{eqnarray}

RR$_i$ and ARR will serve as the evaluation metric for all range partition algorithms throughout this paper. Since ARR only considers user's engagement before the first entity click, it remains a challenge how to measure the performance of a range partition algorithm in the whole session. We leave it for future work. 

\ignore{
 $C_{b}(P, f, i)$ is the sum of the last clicked position in each of the $\kappa$ ranges. When the number of relevant entities increases, the effectiveness of this ranking measurement decreases, because $C_{b}(P, f, i)$ would less rely on $P$. In an extreme case where this number is far greater than $k$, relevant entities are prone to appear in every single range, regardless of the partition $P$. In that case, $C_{b}(P, f, i)$ becomes the position of the last relevant entity in $f$: a constant with respect to $P$. In order to avoid this influence, the number of relevant documents should be controlled and small. A safe strategy is to set the number of relevant document to one, that is, $C_{b}(P, f, i)$ is the minimum browsing cost for the first entity click. \\
}

\ignore{
\textbf{Definition 2. (Normalized Minimum Reranking Cost with Partition $P$ and Ranking Function $f$)}. For each query $q^i$, we can divide $C_{b}(P, f, i)$ by $|E_i|$, i.e., the total number of entities in $E_i$, so it is upper bounded by one. We call this value the \emph{normalized minimum reranking cost with $P$ and $f$}, denoted as $C_{nb}(P, f, i)$. \\
#
Besides the aforementioned relevant document number, we also need to control $k$. As $k$ increases, each range becomes smaller, so $C_{b}(P, f, i)$ and $C_{nb}(P, f, i)$ is always smaller for larger $k$. Finally, we define our problem in Definition 3.\\
#
\textbf{Definition 3. ($k$-Range Partition Cost with Ranking Function $f$)} Given input $\mathcal{Q} = \{q^1, \cdots, q^n\}$, $\mathcal{E} = \{E^1,\allowbreak \cdots, E^n\}$ with fixed number of relevant entities for each $i$, and other useful features (e.g., user profile), a partition algorithm $\mathcal{P}$ (for a numerical facet) generates a mapping from each $i$ to a $k$-range partition: $\mathcal{P}$: $i\rightarrow P^i=\{p_1^i, \cdots, p_{k-1}^i\}$. Algorithm $\mathcal{P}$'s cost is defined as:
\begin{eqnarray}
C_{b}(\mathcal{P},f) &=& \frac{1}{n}\sum_{i=1}^n C_{b}(P^i, f, i)\label{eq:mrc}\\
C_{nb}(\mathcal{P},f) &=& \frac{1}{n}\sum_{i=1}^n C_{nb}(P^i, f, i)\label{eq:nmrc}
\end{eqnarray}
}



\pdfoutput=1

\section{Methods}
\label{sec:approach}

In Section~\ref{sec:intro}, we discuss the quantile method, which partitions $E^i$ into $k$ equal sized ranges. This approach is also used in database system for observing underlying data distribution or data compression (where it is called \emph{equi-depth binning}~\cite{conf/sigmod/MuralikrishnaD88}). Figure~\ref{fig:intro_example_3} shows that the quantile method performs reasonably well. However, quantile method is a simple, rule-based method without leveraging extra information. Suppose we are allowed to use any information we can collect, can we do better than quantile method? 

An idea is to collect another search log for training, since it can help us make better estimation on the testing (evaluation) data. In this section, we propose two methods to leverage the training data. 

\subsection{First Method: Dynamic Programming}
\label{sec:firstmethod}
Since we have defined ARR (Equation~\ref{eq:arr}) as our evaluation metric and the smaller the better, our range partition algorithm should try to minimize ARR and RR$_i$. Imagine if the clicked entity $e^i$ was known, minimizing RR$_i$ means we should make one range only contain $e^i$ itself. RR$_i$ in this imaginary scenario is equal to 1. In reality, although the clicked entity is not known, we can estimate the click probability using the extra search log (i.e., training data). Denote the estimated click probability on entity $e$ as $p(e)$ (so that $\sum_{e\in E^i} p(e) = 1$). Then the expected RR$_i$ for $S=(s_1, \cdots, s_{k-1})$ is:

\vspace{-0.09in}
\begin{eqnarray}
\mathbb{E}_{S}[RR_i] &=& \sum_{e\in E^i} p(e) \times \rrank(e, E^i, S)\label{eq:object}
\end{eqnarray}
\vspace{-0.09in}

So our first method is: for each query $q^i$, to suggest $S^i=\argmin_{S\in \mathbb{R}^{k-1}} \mathbb{E}_{S}[RR_i]$.  

To minimize Equation~\ref{eq:object}, first notice that although $\mathbb{R}^{k-1}$ is continuous, we actually only have to search for $S$ in a discrete subspace of $\mathbb{R}^{k-1}$. The reason is explained in the following example. Suppose $E^i$ only contains three entities (ordered by rank) $e_1,e_2$ and $e_3$. $v(e_1)=100; v(e_2)=200, v(e_3)=300$; estimated probabilities are $p(e_1) = 0.4,\allowbreak p(e_2) = 0.3, \allowbreak p(e_3) = 0.3$; finally, $k=2$, so $S=(s_1)$. Originally, $s_1$ can be any float $\in (100, 300]$ (if $s_1\leq 100$ or $s_1 > 300$, result only contains one range). However, notice objective function (Equation~\ref{eq:object}) stays the same for all $s_1\in (200, 300]$, also for all $s_1\in (100, 200]$. So we only have to pick $a\in(100, 200]$, and $b\in(200, 300]$ and compare the objective function with $S=(a)$ and $S=(b)$. We pick the mid point for convenience, i.e., $a=150$ and $b=250$.

From example above, we can see that in general, minimizing Equation~\ref{eq:object} subject to $S\in \mathbb{R}^{k-1}$ is equal to the combinatorial optimization problem of selecting $k-1$ numbers from $|E^i| - 1$ mid points so that their combined $S$ minimizes the objective function. We can, of course, use brute-force search, but the time cost would be $O({|E^i| - 1 \choose{k-1}} \allowbreak+ |E^i|^3\log{|E^i|})$, where the extra $|E^i|^3\log{|E^i|}$ is for sorting and pre-computing $\rrank(e, E^i, S)$ for each $e$ in each possible range. When $|E^i|$ is large, this time cost is undesirable. However, this problem has a $O(k|E^i|^2 + |E^i|^3\log{|E^i|})$ time solution using dynamic programming. This is because objective function can be rewritten as the sum of $k$ parts, the $k$-th part is independent from previous $k-1$ parts (for proof of this, see Appendix A).

One may wonder why we do not use greedy algorithm here. There are two reasons: first, greedy algorithm generally leads to sub-optimal solutions\footnote{An example: suppose $E^i$ contains four entities (ordered by rank) $e_1, e_2, e_3$ and $e_4$. $v(e_1)=400, v(e_2)=100, v(e_3)=200, v(e_4)=300$, $p(e_1)=p(e_2)=0.2,p(e_3)=p(e_4)=0.3, k=3$. Optimal solution is 1.2 but greedy algorithm's solution is 1.3.}; second, the computational cost of greedy algorithm is $O(k|E^i| + |E^i|^3\log{|E^i|})$, which remains large since it still has to compute ranks of each entity in each possible range. 

\subsection{A Second Look: Parameterization}
\label{sec:percentage}

In Section~\ref{sec:firstmethod}, we propose to suggest $S^i$ that optimizes the expected RR$_i$ for each time $i$. Yet with access to both training and testing data, we have a second thought: can we build a machine learning model to study this problem? 

Take linear regression as an example. Given training data $\{\mathbf{x}^i, y^i\}, i=1,\cdots, n$, it defines parameter $w$ and $b$, finds $w$ and $b$ that minimize the square loss on training data, and applies them on the testing data. In our problem, can we define a set of parameters, model ARR as a function of the parameters, find parameters that minimize ARR on training data, which could then be applied on testing data?

At the first sight, there does not seem to exist a very straightforward solution to the parameterization. One may think $S=(s_1, \cdots, s_{k-1})$ can be the parameters. However, we have discussed in Section~\ref{sec:intro} that it is not a good strategy to use fixed ranges for different queries. On the other hand, we learned that the quantile method performs reasonably well. This sheds light on how we can define the parameters: using the \emph{relative ratio} representation of $S$, i.e., $R=(r_1, \cdots, r_{k-1})\in (0, 1)^{k-1}$ where $r_1 < \cdots < r_{k-1}, r_0=0, r_k = 1$. Given the search results $E^i$, for any $R$, we can find the partition $S$ for $E^i$ so the ratio of number of entities in range $[s_{j-1}, s_j)$ most closest approximates, if not exactly equal to $r_j - r_{j-1}$:

\vspace{-0.2in}
\begin{eqnarray*}
\Delta r_j\coloneqq r _j - r_{j-1} \approx \frac{|\{e\in E^i| v(e)\in [s_{j-1}, s_j)\}|}{|E^i|}
\end{eqnarray*}

The $R$ for quantile method is $(1/k, \cdots, k-1/k)$. With this representation, any $R$ corresponds to one point $(\Delta r_1, \cdots, \allowbreak \Delta r_k)$ in the simplex $\Delta^k$. 

So we want to ask: among all points in $\Delta^k$, does quantile method generate the best ARR on testing data? If not, can we achieve better ARR on testing data by finding parameter $R$ that minimizes the ARR in training data? In this section we study how to optimize ARR with respect to $R$. 

\subsubsection{Optimizing ARR with Respect to $R$}
\label{sec:opt_r}
It is difficult to directly optimize ARR, because same as many evaluation metrics in IR (e.g., NDCG\cite{DBLP:conf/nips/ValizadeganJZM09}, MAP\cite{yue2007support}), ARR is a non-smooth objective function with respect to parameter $R$. Indeed, if the relevant entity is near the boundary, and we change $R$ with a small enough value $\epsilon \rightarrow 0$, relevant entity would jump from one range to another, so RR$_i$ would also jump and as a result, ARR cannot stay continuous. An example: suppose $E^i$ only contains three entities (ordered by rank): $e_1, e_2$ and $e_3$. $v(e_1)=100, v(e_2)=200, v(e_3)=300$; relevant entity is $e_2$ and $k=2$. If we change $R=[0.66]$ to $R’=[0.67]$, the partition would jump from \{\{$e_1$\}, \{$e_2$,$e_3$\}\} to \{\{$e_1$,$e_2$\}, \{$e_3$\}\}, and RR$_i$ would jump from 1 to 2. 

\textbf{Non-smooth optimization}. In order to optimize the non-smooth ARR, first notice that ARR can be non-smooth everywhere, instead of only at a few points\footnote{Therefore our optimization cannot be solved in the same as Lasso~\cite{tibshirani1996regression} which uses sub-gradient descent.}. There exist a few derivative-free algorithms for solving optimization problem in this case. Two of them are Powell's conjugate direction method~\cite{brent73} and Nelder-Mead simplex method~\cite{Nelder1965}, we will discuss more about this topic in Section~\ref{sec:exp}.

\textbf{Time complexity to directly optimize ARR}. Time complexity of directly optimizing ARR with the above non-smooth optimization algorithms is \emph{at least} $O(N_{eval} T_1)$, where $T_1$ is the average time cost to compute ARR on one specific point, and $N_{eval}$ is the number of such points we have to compute (number of function evaluations). In other words, $N_{eval}$ depends on the efficiency of non-smooth optimization algorithm, and $T_1$ depends on the size of the data. We can observe from Equation~\ref{eq:arr} that $T_1 = O(n m\log{m})$, where $n$ is the number of queries in the training data, and $m$ is the average number of retrieved entities $|E^i|$ for each query $q^i$. This is because whenever the optimization algorithm goes to a new point $R$, we have to recompute the ARR from scratch. To explain in more detail: whenever we are at a new point $R$, every RR$_i$ in Equation~\ref{eq:arr} may have changed (as we discussed above, a small enough change in $R$ can lead to a significant change in RR$_i$), so we have to recompute the RR$_i$ in every single query; every such recomputation takes $O(m\log{m})$, which is for sorting entities in the range that contains relevant entity to compute its refined rank. 

In summary, the time complexity for any optimization algorithm to directly optimize ARR is $O(N_{eval}nm\log{m})$. In real world search engines, both $m$ and $n$ can be very large. On the other hand, we are not aware of theoretical estimation on $N_{eval}$, but previous work has provided empirical results. Table 1 to 3 of \cite{journals/coap/GaoH12} show examples of $N_{eval}$ in Nelder-Mead, and Table 2 of \cite{AROUXET2011} shows examples of $N_{eval}$ in Powell's method. Empirically, $N_{eval}$ for lower dimensional problems ($k$ ranges from 2 to 10, which is the case for numerical range partition) usually ranges from 100 to 1,500. 

\subsubsection{Optimizing the Surrogate Objective Function}
\label{sec:surrogate}

As discussed in Section~\ref{sec:opt_r}, the algorithm for directly optimizing ARR takes $O(N_{eval}nm\log{m})$, which is time consuming when $N_{eval}, n, m$ are all very large. In this section, we propose a three-step process that turns ARR into a surrogate objective function. We propose to optimize the surrogate function instead of directly optimizing ARR, so that time cost is significantly reduced. 

\textbf{Step 1: Normalization}. First, for each query $q^i$, we normalize RR$_i$ by the total number of retrieved entities $E^i$:

\vspace{-0.1in}
\begin{eqnarray*}
\overline{RR}_i = \frac{RR_i}{|E^i|} = \frac{\rrank(e^i, E^i, R)}{|E^i|}
\end{eqnarray*}
\vspace{-0.01in}

$\rrank(e^i, E^i, R)$ is the same as $\rrank\allowbreak(e^i, E^i, S)$ where $S$ are the separating values closest to $R$ (see beginning of Section~\ref{sec:percentage}). 

\textbf{Step 2: Upper bound}. By definition (Section~\ref{sec:metric}), $\rrank(e^i, E^i, R)$ is bounded by the total number of entities in the unique range that contains relevant entity $e^i$. Denote this range as $[s_{j_i}, s_{j_i + 1})$:

\vspace{-0.1in}
\begin{eqnarray}
\overline{RR}_i & \leq & \frac{|\{e\in E^i|v(e)\in [s_{j_i}, s_{j_i + 1})\}|}{|E^i|}\label{eq:upperbound}
\end{eqnarray}
\vspace{-0.05in}

\textbf{Step 3: Limit approaching infinity}. Notice as $|E^i|$ goes to infinity, the R.H.S. of Inequality \ref{eq:upperbound} approaches $\Delta r_{j+1} = r_{j+1} - r_j$ (see beginning of Section~\ref{sec:percentage}). If we denote $z^i$ as the ratio of number of entities smaller than or equal to $v(e^i)$\footnote{For example: suppose $E^i$ only contains four entities (ordered by rank): $e_1,e_2,e_3$ and $e_4$. $v(e_1)=100, v(e_2)=300, v(e_3)=200, v(e_4)=400$; relevant entity is $e_2$. In this example, $z^i=\frac{3}{4}$.}, this limit is rewritten as:

\vspace{-0.2in}
\begin{eqnarray*}
C^i(R) \coloneqq \Delta r_{j_i + 1} = \sum_{j=1}^k \mathbbm{1}[r_{j-1}\leq z^i \leq r_j] \times \Delta r_j
\end{eqnarray*}
\vspace{-0.1in}

The averaged limit over $i=1\cdots, n$ is defined as $C_n(R)$:
\vspace{-0.1in}
\begin{eqnarray}
C_n(R) &=& \frac{1}{n}\sum_{i=1}^n C^i(R)\nonumber\\
&=& \sum_{j=1}^k \Delta r_j \times (F_n(r_j) - F_n(r_{j-1}))\label{eq:cr}
\end{eqnarray} 
\vspace{-0.1in}

Where $F_n(r) = \frac{1}{n}\sum_{i=1}^n \mathbbm{1}[z^i < r]$ for $r\in [0, 1]$ is exactly equal to the empirical conditional distribution function (CDF) of $z^i$. Second equation in (\ref{eq:cr}) follows from simple math. So instead of directly optimizing ARR, we propose to optimize $C_n(R)$ instead.

\textbf{Time complexity to optimize $C_n(R)$}. We can see the time cost for optimizing $C_n(R)$ is largely reduced compared with ARR. This is because the empirical CDF $F_n(r)$ can be first computed and cached using Algorithm~\ref{algo}. After $F_n(r)$ is cached, at any new point $R$ where the non-smooth optimization algorithm needs to re-compute $C_n(R)$, it only have to obtain the cached $F_n(r)$ from $X_{sorted}$ and $Y$ (output from Algorithm~\ref{algo}) for $r = r_1, \cdots, r_{k-1}$ then apply Equation~\ref{eq:cr}. To obtain cached $F_n(r)$, we first use binary search on $X_{sorted}$ to find the index $i$ of $r$, then return $Y[i]$ as $F_n(r)$. Therefore, time complexity for each of the $N_{eval}$ function evaluation is reduced to $O(k\log n_0)$. 

Time costs for caching $F_n(r)$ are listed in Algorithm~\ref{algo}. In summary, the total time complexity for caching + optimizing $C_n(R)$ is $O(nm + n_0\log{n_0} + n \log{n} + n_0\log{n} + N_{eval}k\log{n_0})$. $n_0$ is the number of unique $r_j's$ in the log, so $n_0 < |X_{ct}|m < nm$. 

\vspace{-0.05in}
\begin{algorithm}
\caption{Caching Empirical CDF $F_n(r)$\label{algo}}
\DontPrintSemicolon
  $X_{ct}\leftarrow \emptyset$; \tcp*{Set of unique $|E^i|$}
  $X\leftarrow \emptyset$; \tcp*{Set of unique $r_j$'s}
  $Y\leftarrow []$; \tcp*{$F_n(r_j)$ values of all unique $r_j$'s}
  $Z\leftarrow []$; \tcp*{All $z^i$'s}
  \For{$i=1,\cdots, n$}{
  \If{$|E^i|\not\in X_{ct}$}{
    $X_{ct}\leftarrow X_{ct} \cup \{|E^i|$\};\;
    \For{$j=1,\cdots, |E^i| - 1$}{
      $X\leftarrow X\cup \{\frac{j}{|E^i|}$\};
    }
  }
      $count\leftarrow 0$;\;
    \For{$e\in E^i$}{
      \If{$v(e) \leq v(e^i)$}{
        $count \leftarrow count + 1$;\tcp*{$O(nm)$}
      }
    }
    $z^i\leftarrow count/|E^i|$;\;
    Append $z^i$ to the end of $Z$;
  }
  $n_0\leftarrow |X|$;\;
  $X_{sorted}\leftarrow sort(X)$;\tcp*{$O(n_0\log{(n_0)})$}
  $Z_{sorted}\leftarrow sort(Z)$;\tcp*{$O(n\log{n})$}
  \For{$i=1,\cdots, |X_{sorted}|$}{
      $x\leftarrow X_{sorted}[i]$;\;
      $Pos\leftarrow BinarySearch(Z_{sorted}, x)$;\tcp*{$O(n_0\log{n})$}
      $y\leftarrow Pos/n$;\;
      Append $y$ to the end of $Y$;\;
  }
  \Return $X_{sorted}$ and $Y$;
\end{algorithm}
\vspace{-0.2in}

\subsubsection{Bounds on $C_n(R)$}
The Dvoretzky-Kiefer-Wolfowitz inequality~\cite{Karimzadehgan:2010:ETI:1871437.1871631} bounds the probability that the empirical CDF $F_n$ differs from the true distribution $F$. Following the DKW inequality, we are able to prove a few bounds on $C_n(R)$. These bounds provide useful insights on the convergence rate and sample complexity of $C_n(R)$ on large scale datasets. We show them in Appendix B.

\subsection{Learning to Partition with Regression Tree}
\label{sec:tree}
In Section~\ref{sec:percentage} we propose to optimize $C_n(R)$ subject to the ratio parameter $R$, and apply it to the testing data. This means all queries in testing data shares the same $R$. If they can have different $R$'s, can we further improve the results?

To differentiate each query, we define a feature vector $\mathbf{x}^i\in \mathbb{R}^d$ for query $q^i$. For example, $\mathbf{x}^i$ can be $q^i$'s low dimensional representation using the latent semantic analysis (LSA). A heuristic solution, for example, is to replace $R$ with $R^i = \beta^T \mathbf{x}^i$ in each query, and optimize $C_n$ subject to $\beta^T$. However, $C_n$ defined this way is much harder to optimize, because $\Delta r_j$ is now different for each query, so $F_n(r)$ can no longer be pre-computed and cached.

This observation implies that we should try to make each $R^i$ shared by at least a significant number of queries. The best machine learning method under this setting (that we are aware of) is the regression tree (CART \cite{BreimanEtAl:84}). In a regression tree, all queries inside each leaf node $t$ share the same parameter $R_t$. 

Training of a regression tree would recursively split examples in the current node. In each node, it chooses the dimension $j\in [d]$ and the threshold $\theta$ so that splitting by whether $\mathbf{x}^i_j > \theta$ minimizes the sum of mean square error (MSE) on each side. The overall goal of regression tree is to minimize the square error on training data. On the other hand, our goal is to minimize the ARR on training data, and because ARR is hard to compute, we minimize $C_n(R)$ instead (Section~\ref{sec:surrogate}). Therefore, we can build a regression tree for our problem where the splitting criterion at each node is to select $j\in [d]$ and $\theta$ to minimize the sum of minimum $C_n(R)$ on each side. 

\vspace{-0.05in}
\begin{itemize}
\item \textbf{Splitting criterion 1}. Select dimension and separating value that minimizes $C_n$ (Equation~(\ref{eq:cr})); 
\end{itemize}
\vspace{-0.05in}

However, it is interesting to observe how minimizing MSE resembles minimizing $C_n$. Imagine two different splits on the same data. Suppose that with one split, data is perfectly separated into two clusters; with the other split, however, data is still well mixed. The former one would have smaller MSE. It would also have smaller $C_n$, since $R$ in each cluster is highly fitted in a small region. Therefore, we propose to use MSE as an alternative splitting criterion:

\vspace{-0.05in}
\begin{itemize}
\item \textbf{Splitting criterion 2}. Select dimension and separating value that minimizes the mean square error; 
\end{itemize}
\vspace{-0.05in}

Criterion 2 does not compute the parameter $R$, so after the tree is constructed, we need extra time to compute $R_t$ for each node $t$. But even so, Criterion 2 is orders of magnitude faster than Criterion 1. This is because, on the one hand, while Criterion 1 needs to reconstruct a new tree for every $k$, criterion 2 only needs to build one tree the whole time. On the other hand, time cost of criterion 2 in constructing each tree is significantly less than criterion 1, because computing MSE is much faster than minimizing $C_n$. 

An important step in regression tree \cite{BreimanEtAl:84} is the minimal cost-complexity pruning. First, a full (overfitting) tree is grown, then the algorithm goes through 5 fold cross validation to select the optimal pruning for the fully grown tree. We apply the same pruning strategy for Criterion 1 and 2, where we use the 0.5 SE rule to select the optimal tree. 

\subsection{Testing Time and Rounding}
\label{sec:time}
\textbf{Testing complexty}. For each $q^i$, testing time for our first method (Section~\ref{sec:firstmethod}) is $O(k|E^i|^2 + |E^i|^3\log{|E^i|})$ Our second method (both Section~\ref{sec:opt_r} and Section~\ref{sec:tree}) takes constant time to generate $R^i$, but the $R^i$ still needs to be converted back to $S^i$. There are two ways to do this: first, sort $E^i$ by $v(e)$, which takes $O(|E^i|\log{|E^i|})$; second, apply the k-th smallest element algorithm\footnote{e.g., quickselect \url{https://en.wikipedia.org/wiki/Quickselect}}, which takes $O(k|E^i|)$. When $|E^i|$ is large, this step can also be time consuming. However, we have to scan $E^i$ for at least one time anyway. This is because after $S^i$ is generated, for all $e\in E^i$ we need to find the range that contains it. So second method does not increase time complexity with respect to $|E^i|$. 

\textbf{Rounding}. To better user experience, we need to generate easy-to-read ranges, therefore we may need to round the floating numbers in $S^i$. Rounding precision depends on the application scenario. For price of products, users may be expecting more friendly designs, thus they may prefer `Below 150' to `Below 149.7'. In other applications such as distance, users may accept higher precision such as `Below 11.7 miles'. The rounding precision can also be tuned as a parameter. 

\pdfoutput=1

\begin{table*}[h]
\begin{center}
    \begin{tabular}{|>{\centering\arraybackslash}p{1cm} | >{\centering\arraybackslash}p{1cm} | >{\centering\arraybackslash}p{1cm} | >{\centering\arraybackslash}p{1cm} | >{\centering\arraybackslash}p{1cm} | >{\centering\arraybackslash}p{1cm} || >{\centering\arraybackslash}p{1cm} | >{\centering\arraybackslash}p{1cm} | >{\centering\arraybackslash}p{1cm} | >{\centering\arraybackslash}p{1cm} | >{\centering\arraybackslash}p{1cm} | >{\centering\arraybackslash}p{1cm} | }
    \hline
    \multicolumn{2}{|l|}{\multirow{3}{*}{}} & \multirow{2}{*}{\texttt{quant.}} & \multirow{2}{*}{\texttt{dp}} & \multirow{2}{*}{\texttt{powell}} & \multirow{2}{*}{\texttt{tree}} & \multicolumn{2}{c|}{\texttt{tree} vs. \texttt{dp}} & \multicolumn{2}{c|}{\texttt{tree} vs. \texttt{quant.}} & \multicolumn{2}{c|}{\texttt{dp} vs. \texttt{quant.}}\\  \cline{7-12}
 \multicolumn{2}{|c|}{} & & & & & $p$ & $t$ & $p$ & $t$ & $p$& $t$\\ \hline
 \multirow{5}{*}{\specialcell{Laptop}} & $k=2$ & 33.27 & 30.15 & 31.63  & \textbf{28.00} & 0.32 & -0.98 & 9e-3 & -1.45 & 0.15 & -1.45\\ \cline{2-12}
& $k=3$ & 22.07 & 21.22 & 19.95 & \textbf{17.62} & 0.03 & -2.18 & 5e-4 & -3.50 & 0.61 & -0.50\\ \cline{2-12}
& $k=4$ & 16.76 & 16.47 & 15.28 & \textbf{13.29} & 0.02 & -2.23 & 3e-4 & -3.63 & 0.83 & -0.20\\ \cline{2-12}
& $k=5$ & 13.55 & 13.43 & 11.94 & \textbf{10.72} & 0.04 & -2.05 & 3e-4 & -3.65 & 0.92 & -0.09\\ \cline{2-12}
& $k=6$ & 11.33 & 11.03 & 10.15 & \textbf{9.03} & 0.04 & -2.02 & 2e-4 & -3.69 & 0.76 & -0.29\\ \hline\hline
\multirow{5}{*}{\specialcell{TV}} & $k=2$ & 31.85 & 30.99 & 31.73 & \textbf{30.78} & 0.89 & -0.12 & 0.49 & -0.68 & 0.60 & -0.52\\ \cline{2-12}
& $k=3$ & 21.30 & 20.88 & 21.43 & \textbf{20.75} & 0.89 & -0.12 & 0.60 & -0.51 & 0.69 & -0.38 \\ \cline{2-12}
& $k=4$ & 16.19 & 15.95 & 16.30 & \textbf{15.57} & 0.63 & -0.47 & 0.43 & -0.78 & 0.76 & -0.29 \\ \cline{2-12}
& $k=5$ & 13.08 & 12.83 & 13.18 & \textbf{12.62} & 0.75 & -0.31 & 0.47 & -0.72 & 0.70 & -0.37 \\ \cline{2-12}
& $k=6$ & 10.95 & 10.64  & 10.98 & \textbf{10.48} & 0.76 & -0.30 & 0.37 & -0.89 & 0.57 & -0.55 \\ \hline
    \end{tabular}
\end{center}
\vspace{-0.15in}
\caption{Comparative study on the ARR of four methods. The ARR metric can be interpreted in this way: when the number of partitioned ranges is 6, users needs to read 11.33 products in average with \texttt{quantile} method; while she only needs to read 9.03 products in average with \texttt{tree} method. \texttt{dp}, \texttt{powell} and \texttt{tree} uses the same amount of training data for fair comparison.\label{tab:results}}
\end{table*}

\section{Experiments}
\label{sec:exp}

In this section, we conduct comparative experiments on the quantile method and our two methods to answer the question in Section~\ref{sec:intro} and Section~\ref{sec:approach}, i.e., can we leverage previous search logs to improve the results on test collection?

\subsection{Dataset}
\label{sec:data}

Since no existing work has studied our problem setting (Section~\ref{sec:problem}), we have to construct our own dataset. We collect a two-month search log from \url{www.walmart.com} between 2015/10/22 and 2015/12/22. Since the size of the entire log is intractable on a single machine, we only keep the data from two categories: `Laptop' and `TV', because they are among the categories with the most traffic. Our data contains multiple numerical facets, e.g., screen size and memory capacity. We select the price facet for experiment, because most product (larger than 90\%) contains this facet. Although price can vary from time to time, we assume it is fixed within a short period of time, so each product in our data can only contain one price. 

For each category, we separate the earlier 70\% as training data and latter 30\% as testing data (according to timestamps). After the separation, Laptop contains 2,279 training queries and 491 testing queries, TV contains 4,026 training queries and 856 testing queries. Data structure under each query is the same as the input described in Section~\ref{sec:problem}, plus the ground truth of which entity is clicked (relevant). 

\subsection{Experimental Results}
We compare ARR generated by four methods on testing data: \texttt{quantile}: for each query, quantile method generates $k$ ranges so each range contains the same number of products; \texttt{dp}: for each query, our first method (Section~\ref{sec:firstmethod}) generates $k$ ranges which optimize expected RR$_i$ (Equation~\ref{eq:object}) using DP; \texttt{powell}: (Section~\ref{sec:percentage}) first use Powell's method to find $R$ by optimizing $C_n(R)$ (Equation~\ref{eq:cr}) on training data, then apply the same $R$ to all queries on testing data; \texttt{tree}: find different $R$'s using regression tree (Section~\ref{sec:tree}) and apply the tree to all queries on testing data. 

Of the four methods, \texttt{quantile} does not leverage training data; we use all training data to estimate $p(e)$ for \texttt{dp} (which we discuss in details in Section~\ref{sec:comparepe}), so \texttt{dp}, \texttt{powell} and \texttt{tree} use the same amount of training data.

\subsubsection{Overall Comparative Study}
Table~\ref{tab:results} shows the ARR of the four methods. For every method we report the best tuned ARR by varying its parameters. We can see that the overall performance of \texttt{tree} is the best among all; \texttt{powell} and \texttt{dp} are next, with \texttt{powell} slightly better in Laptop and \texttt{dp} slightly better in TV; \texttt{quantile} has the worst performance in Laptop, and similar performance as \texttt{powell} in TV. On the other hand, if we vertically compare Laptop vs. TV in each method, we can see that \texttt{quantile} and \texttt{dp} are slightly better in TV than Laptop, while \texttt{powell} and \texttt{tree} are the opposite. 

We run T-test between each pair of methods in \texttt{quantile}, \texttt{dp} and \texttt{tree}. We skip T-test on \texttt{powell} because \texttt{tree} generalizes \texttt{powell}, and Table~\ref{tab:results} shows \texttt{tree} always outperforms \texttt{powell}. From Table~\ref{tab:results} we can see that T-test results are different in Laptop and TV. For Laptop, \texttt{tree} significantly ourperforms \texttt{quantile} and \texttt{dp} (except for \texttt{tree} vs. \texttt{dp} when $k=2$, which may be because performance of parameterized method is hurted when degree of freedom = 1); for TV, however, T-test results are not significant; also, \texttt{dp} vs. \texttt{quantile} are not significant.

These analyses indicate \texttt{tree} and \texttt{powell} perform especially well on Laptop data. So what causes the difference between TV and Laptop?

\begin{figure}[h]
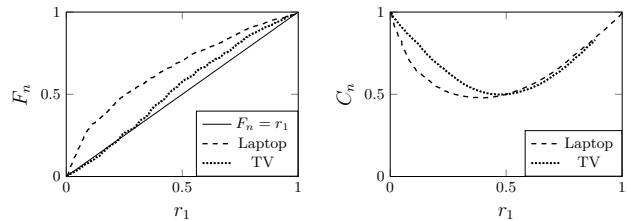

\centering
\vspace{-0.2in}
\subfloat{
\input{plot_cdf.tikz}
}
\subfloat{
\input{plot_c.tikz}
}
\vspace{-0.1in}
\caption{$F_n$ and $C_n$ for Laptop and TV when $k=2$\label{fig:laptoptv}}
\end{figure}

\begin{table}[ht]
\vspace{-0.05in}
\centering
\begin{tabular}{|c|c|c|c|}
\hline
   & $k=2$ & $k=3$ & $k=4$ \\
\hline
\texttt{exhaustive} & 31.72 & 21.27 & 16.14\\ \hline
\texttt{quantile} & 31.85 & 21.30 & 16.19\\ \hline
\end{tabular}
\caption{Optimal ARR vs. \texttt{quantile}'s ARR for `TV'\label{tab:opt}}
\vspace{-0.05in}
\end{table}

To answer this question, we need to find out how \texttt{powell} and \texttt{tree} really works. Recall that \texttt{powell} optimizes $C_n(R)$, which is computed from $F_n(r)$ (Equation~\ref{eq:cr}). When $k=2$, that is, $R=(r_1)$, we are able to plot $C_n(R)$ and $F_n(r)$ as a function of $r_1$. We show the two plots in Figure~\ref{fig:laptoptv}. From Figure~\ref{fig:laptoptv} we can see: $F_n$ of TV is very close to linear, and (consequently) $C_n$ of TV is very close to a quadratic function whose minimum point is $r_1=0.5$ (Indeed, by plugging $F_n(r_1) = r_1$ into Equation~\ref{eq:cr} we get $C_n(r_1) = 2r_1^2 - 2r_1 + 1$). For general $k$, the minimum point $R$ found by these algorithms is almost equal to \texttt{quantile} method. In other words, \texttt{quantile} almost reaches the optimal $R$ on training data in terms of $C_n(R)$.

But our final goal is to optimize ARR \emph{on testing data}. Has \texttt{quantile} method also reached the optimal $R$ on testing data in terms of ARR? To find out the true optimal $R$ on testing data, we perform grid search. We exhaustively enumerate $r_j (j=1,\cdots, k-1$) over all candidate values (i.e., $X_{sorted}$ in Algorithm~\ref{algo}); at each point, we evaluate the true ARR on testing data, and return the minimum value we find. Time complexity of this exhaustive search is $O({{n_0}\choose{k-1}})$. When $k>4$, it becomes intractable. We thus only compute the results for $k\leq 4$\footnote{Although it seems we can replace exhaustive search with Powell's method, which is efficient thus can be applied to $k > 4$; notice Powell's method can not guarantee finding global optimal like exhaustive search.} and show them in Table~\ref{tab:opt} (\texttt{exhaustive}), compared with ARR of \texttt{quantile} method. From Table~\ref{tab:opt} we can see that \texttt{quantile} method indeed almost achieves optimal. So it is difficult for \texttt{tree} and \texttt{powell} to outperform \texttt{quantile}. 

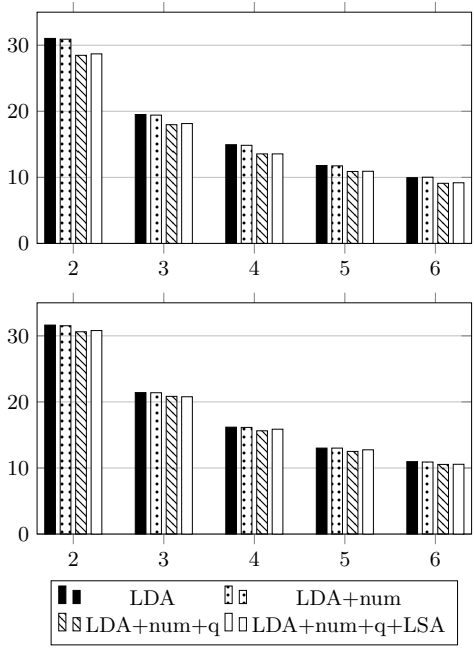
\begin{figure}[h]
\centering
\subfloat{
\begin{tikzpicture} [scale=0.9]
\begin{groupplot}[group style={group size= 1 by 2},height=5cm,width=8cm]
\nextgroupplot[ybar,symbolic x coords={2, 3, 4, 5, 6},  ymin=0, ymax=35,ymajorgrids = true,bar width = 4.5,xtick=data]
\addplot[fill=black,draw=black] 
coordinates {(2,31.03)	(3,19.51)	(4,14.94)	(5,11.79)	(6,9.95)};
\addplot[fill=white,draw=black,pattern=dots] 
coordinates {(2,30.91)	(3,19.42)	(4,14.85)	(5,11.74)	(6,10.02)};
\addplot[fill=white,draw=black,pattern=north west lines] coordinates {(2,28.47)	(3,17.97)	(4,13.55)	(5,10.89)	(6,9.1)};
\addplot[fill=white] coordinates {(2,28.69)	(3,18.14)	(4,13.55)	(5,10.92)	(6,9.18)};
\end{groupplot}
\end{tikzpicture}
}\\
\vspace{-0.1in}
\subfloat{
\begin{tikzpicture} [scale=0.9]
\begin{groupplot}[group style={group size= 1 by 2},height=5cm,width=8cm]
\nextgroupplot[ybar,symbolic x coords={2, 3, 4, 5, 6},legend style={at={(0.5,-0.2)},anchor=north, ymin=0, ymax=35,legend columns=2},  ymajorgrids = true,bar width = 4.5,xtick=data]
\addplot[fill=black,draw=black] 
coordinates {(2,31.63)	(3,21.41)	(4,16.18)	(5,13.01)	(6,10.96)};
\addplot[fill=white,draw=black,pattern=dots] 
coordinates {(2,31.53)	(3,21.39)	(4,16.14)	(5,13.01)	(6,10.9)};
\addplot[fill=white,draw=black,pattern=north west lines] coordinates {(2,30.62)	(3,20.84)	(4,15.63)	(5,12.51)	(6,10.54)};
\addplot[fill=white] coordinates {(2,30.8)	(3,20.78)	(4,15.87)	(5,12.75)	(6,10.56)};
\addlegendentry{LDA}
\addlegendentry{LDA+num}
\addlegendentry{LDA+num+q}
\addlegendentry{LDA+num+q+LSA}
\end{groupplot}
\end{tikzpicture}
}

\caption{Compare importance of different feature groups: ARR for $k=2,\cdots, 6$. Above: Laptop; below: TV\label{fig:treefeat}}
\vspace{-0.2in}
\end{figure}

\subsubsection{Comparative Study on Non-smooth Optimization Methods}
\label{sec:nonsmooth}

In this section we conduct comparative study on the performance of different non-smooth optimization methods. We study five non-smooth algorithms. Besides the aforementioned 1) \texttt{powell} and 2) \texttt{nelder-mead}, we also study: 3) \texttt{cg}: conjugate gradient method in non-smooth case; 4) \texttt{bfgs}: second order optimization method in non-smooth case; and
5) \texttt{slsqp}: sequential least square programming. For all the five methods we use the implementation in Python library\footnote{\url{https://docs.scipy.org/doc/scipy/reference/generated/scipy.optimize.minimize.html}}. For each algorithm, we run 5 fold cross validation to tune the error tolerance as well as to find a good starting point. We report the performance of each algorithm in Table~\ref{tab:nonsmooth}. Due to space limit and since our goal is comparative study, results in Table~\ref{tab:nonsmooth} is the average over $k=2,\cdots, 6$. To ensure the statistical significance, we randomly restart each algorithm 50 times and report the average (i.e., each number in Table~\ref{tab:nonsmooth} is averaged over 50$\times$ 5 values). 

From Table~\ref{tab:nonsmooth} we can see that the five algorithms have slightly different performances: \texttt{slsqp} has the best performance in Laptop and \texttt{powell} has the best performance in TV. \texttt{powell} and \texttt{nelder-mead} has the largest time cost, while \texttt{bfgs} is the fasted algorithm among all. This can be explained by the fact that \texttt{bfgs} is a second order method, while \texttt{Powell} and \texttt{nelder-mead} does not leverage the gradient information compared with the other three. 
\begin{table}
\centering
\begin{tabular}{|c|c|c|c|c|c|c|}
\hline
  \multicolumn{2}{|c|}{} & \texttt{powell} & \texttt{bfgs}& \texttt{nelder} & \texttt{cg} &\texttt{slsqp}\\
\hline
\multirow{2}{*}{\specialcell{avg\\ARR}} &L & 17.77 &  17.58 &  17.78 & 17.60  & \textbf{17.50} \\ \cline{2-7}
&T & \textbf{18.70} & 18.76 & 18.74& 19.06 & 18.76\\ \hline\hline
\multirow{2}{*}{time}&L & 0.024 & \textbf{0.007}&0.028 & 0.012&0.027\\ \cline{2-7}
&T & 0.022 &\textbf{0.008}&0.026 &0.009 &0.009\\ \hline
\end{tabular}
\caption{Compare different non-smooth optimization methods: averaged ARR and running time over $k=2,\cdots, 6$.\label{tab:nonsmooth}}
\vspace{-0.1in}
\end{table}

\subsubsection{Comparative Study on Regression Tree Features}

Since regression tree method (Section~\ref{sec:tree}) uses feature $\mathbf{x}^i$ for each query $q^i$, in this section, we study the influence from different features. We use three groups of features:

\textbf{Semantic representation for $q^i$}: we use both latent semantic analysis (LSA) and latent Dirichlet allocation (LDA). For each method the dimension is set to 20. 


\textbf{Number of explicitly mentioned facets in $q^i$}: we use Stanford Named Entity Recognizer (NER) to label the explicitly mentioned facets in each query. For example, for query `17 in refurbished laptop', explicitly mentioned facets are screen size=17 and condition=refurbished, so this feature = 2. We manually label 40\% of the queries for training, the rest are computed by the recognizer. Intuition behind this feature is when user mentions more facets, it is more likely she is looking for a higher profiled product;

\textbf{Quartile absolute values of numerical facets in $E^i$}: we use quartile facets, which are absolute values of the 25\%, 50\% and 75\%th smallest facets in $E^i$. Intuition behind this feature is when retrieved products are all very expensive, user may prefer relatively less expensive products in the list;

 We study four combinations of these features\footnote{In this experiment the splitting criterion of regression tree is fixed to criterion 2 and non-smooth optimization method is fixed to Powell's method. }: (1) LDA (dimension=20): using only 20 features from LDA; (2) LDA + num (dimension=21): adding the number of explicitly mentioned facets; (3) LDA + num + q (dimension=24): \allowbreak adding the quartile absolute value features; (4) LDA + num + q + LSA (dimension=44): adding 20 features from LSA. The comparative results of the four groups is shown in Figure~\ref{fig:treefeat}. Figure~\ref{fig:treefeat} shows that quartile absolute value features is most helpful in reducing ARR; number of explicitly mentioned facets does not help a lot; LSA features also do not help ARR, actually hurts ARR in many cases, which can be explained by the fact that we already have LDA features. 

\subsubsection{Comparative Study on Regression Tree Splitting Criterion}
\begin{figure}
\centering
\subfloat{
\begin{tikzpicture} [scale=0.9]
\begin{groupplot}[group style={group size= 1 by 2},height=5cm,width=6.4cm]
\nextgroupplot[ybar,symbolic x coords={2, 3, 4, 5, 6},xtick=data,yticklabels={a, -100\%, -50\%, 0\%, 50\%, 100\%},ylabel={\texttt{nonsquare} $\hspace{5mm}$ \texttt{square}}, ymin=-1,ymax=1,legend style = {at={(1.05, 0.5)}, anchor = west, legend columns =1, draw=none, area legend},  ymajorgrids = true,bar width = 4.5]
\addplot[fill=white,draw=black,pattern=north west lines] 
coordinates {(2,-0.819051682808)  (3,-0.771124916997)  (4,-0.778156432501)  (5,-0.826564394207)  (6,-0.805071372915)};
\addplot[fill=white,draw=black,pattern=dots] 
coordinates {(2,-0.233907427982)  (3,-0.497692772272)  (4,-0.194701640859)  (5,-0.298327841493)  (6,-0.319859994764)};
\addplot[fill=black,draw=black] coordinates {(2,-0.774601624292)  (3,0.713975714954)  (4,0.604112124159)  (5,0.634194817398)  (6,0.72013948444)};
\addlegendentry{\textsf{best}}
\addlegendentry{\textsf{min}}
\addlegendentry{\textsf{full}}
\end{groupplot}
\end{tikzpicture}
}\\
\vspace{-0.1in}
\subfloat{
\begin{tikzpicture} [scale=0.9]
\begin{groupplot}[group style={group size= 1 by 2},height=5cm,width=6.4cm]
\nextgroupplot[ybar,symbolic x coords={2, 3, 4, 5, 6},xtick=data,yticklabels={a, -100\%, -50\%, 0\%, 50\%, 100\%},ylabel={\texttt{nonsquare} $\hspace{5mm}$ \texttt{square}}, ymin=-1,ymax=1,legend style = {at={(1.05, 0.5)}, anchor = west, legend columns =1, draw=none, area legend},  ymajorgrids = true,bar width = 4.5]
\addplot[fill=white,draw=black,pattern=north west lines] 
coordinates {(2,0.914413785446)  (3,-0.930402389178)  (4,-0.943460841988)  (5,0.863505531775)  (6,-0.904933980344)};
\addplot[fill=white,draw=black,pattern=dots] 
coordinates {(2,-0.634162717384)  (3,-0.60961347135)  (4,-0.49887440173)  (5,-0.632160292922)  (6,-0.661732896002)};
\addplot[fill=black,draw=black] coordinates {(2,-0.844457306576)  (3,0.477461547435)  (4,0.926149429859)  (5,0.440359870388)  (6,0.645586370174)};
\addlegendentry{\textsf{best}}
\addlegendentry{\textsf{min}}
\addlegendentry{\textsf{full}}
\end{groupplot}
\end{tikzpicture}
}

\caption{Compare different splitting criteria for regression tree method: $p$-value in T-test between minimizing mean square error (\texttt{square}) and minimizing $C_n$ (\texttt{nonsquare}). Above: Laptop; below: TV\label{fig:split}}
\vspace{-0.1in}
\end{figure}
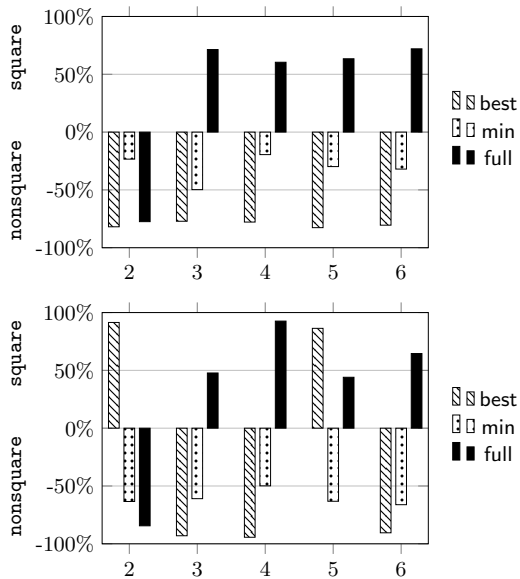

In Section~\ref{sec:tree}, we discuss the usage of two splitting criteria for building the regression tree. Recall the first criterion is to minimize $C_n(R)$ (Equation~\ref{eq:cr}), while the second criterion is to minimize MSE. Therefore, we denote the first criterion as \texttt{nonsquare} and the second criterion as \texttt{square}. In this section, we study the influence of splitting criterion on the performance of regression tree. In order to make a comprehensive comparison, we look into three trees under each criterion: first, fully grown tree without pruning, denoted as \textsf{full}; second, the smallest tree after pruning, which only contains the root node and two leaf nodes, denoted as \textsf{min}; third, the best ARR among all the pruned trees and the fully grown tree, denoted as \textsf{best} \footnote{In this experiment $\mathbf{x}^i$ is fixed to LDA + num + q and non-smooth optimization method is fixed to Powell's method.}. In Figure~\ref{fig:split} we show $p$ values in the T-test results between the two criteria. When criterion 2 is better, we plot the $p$ value in positive (\texttt{square}); otherwise, we plot the $p$ value in negative (\texttt{nonsquare}). 

From Figure~\ref{fig:split} we can see that the difference between the two criteria are basically consistent over $k=2,\cdots, 6$. Although none of the $p$ values is small enough to show statistical significance, we can still observe a few phenomena: first, \textsf{best} of \texttt{nonsquare} is slightly better than \texttt{square}; second, \textsf{min} of \texttt{nonsquare} is more significantly better than \texttt{square}; third, \textsf{full} of \texttt{square} is instead better. These observations can be naturally explained: since the splitting criterion of \texttt{nonsquare} is to optimize $C_n$ which approximates ARR, it is expected to achieve better ARR than \texttt{square}, for the same reason its \textsf{min} should also have better performance. Meanwhile, due to the scarcity of data samples in leaf nodes, \textsf{full} of \texttt{nonsquare} should be more overfitted than \texttt{square}, because it tries to fit ARR in every possible step. 

\subsubsection{Comparative Study on $p(e)$} 
\label{sec:comparepe}

\begin{table}[ht]
\centering
\begin{tabular}{|c|c|c|c|c|c|}
\hline
   & $k=2$ & $k=3$ & $k=4$ & $k=5$ & $k=6$\\
\hline
L & 63.44 & 59.65 & 55.98 & 54.78 & 51.75\\ \hline
T &  61.78& 60.42 & 59.39 & 58.29 & 57.16\\ \hline
\end{tabular}
\caption{ARR using $p(e)\propto 1/rank(e)$\label{tab:dp}}
\vspace{-0.1in}
\end{table}

In this section we study the performance of the DP algorithm using different $p(e)$'s. First, $p(e)$ used in Table~\ref{tab:results} is a combination of the query relevance and the category relevance models:
\begin{eqnarray*}
p(e) &=& \lambda p_q(e) + (1-\lambda) p_{cate}(e)\label{eq:mle}\\
p_{cate}(e) &\propto& \#click(e, cate)\nonumber\\
p_{q}(e) &\propto& \#click(e, q)\nonumber
\end{eqnarray*}
where $\#click(e, cate)$ is the number of clicks on product $e$ under category $cate$; $\#click(e, q)$ is the number of clicks on $e$ under query $q$. These number of clicks are counted from the entire training data (Section~\ref{sec:data}). As a result, DP in Table~\ref{tab:results} uses the same amount of training data as \texttt{tree} and \texttt{powell}. The best tuned parameter $\lambda = 0.5$, which we use in Table~\ref{tab:results}.

Alternatively, $p(e)$ can be estimated from $e$'s rank on \url{www.walmart.com}, i.e., $p(e)\propto 1/rank(e)$. To compare the performance of two methods for estimating $p(e)$, we display the ARR of the second method in Table~\ref{tab:dp}. From Table~\ref{tab:dp} and Table~\ref{tab:results} we can see the first method significantly outperforms the second one, which explains that leveraging training data can help improve the performance of our first method.


\pdfoutput=1

\section{Conclusion}
\label{sec:discussion}
In this paper, we introduce a new problem of numerical facet range partition. We propose evaluation metric ARR based on the browsing cost for user to navigate into relevant entities. We propose two methods that leverages training data, and compare them with the quantile method which does not use training data. Experimental results show that for the TV category, quantile method already achives near-optimal performance; while for Laptop, our second method significantly outperforms quantile method, it even significantly outperforms our first method, which leverages the same amount of training data. Our second method is robust and efficient, so it can be directly applied to any search engine that supports numerical facets.

Future directions include: First, how to generate ranges for interactive search? How to improve partition based on previous user feedback? Second, is there an easily interpretable way of partitioning categorical facets, e.g., brand? Third, how to tune parameter $k$ and rounding precision? 


 \section*{Acknowledgement}

 This work is supported in part by NSF under Grant Numbers CNS-1513939 and CNS-1408944.

\clearpage
\bibliographystyle{abbrv}
\bibliography{facet_partition.bib}

\clearpage
\appendix

\pdfoutput=1

\section{Proof for DP}

\begin{eqnarray*}
&&\rrank(e, E^i, S^i) \\
&=& \sum_{j=1}^k \mathbbm{1}[s_{j-1}\leq v(e) < s_j] \times RRank(e, E^i, [s_{j-1}, s_j))
\end{eqnarray*}
Where $RRank(e, E^i, [s_{j-1}, s_j))$ is the rank of entity $e$ in range $[s_{j-1}, s_j)$. 
\begin{eqnarray*}
&&\Rightarrow \\
&& \mathbb{E}_S[RR_i]\\
&=& \sum_{j=1}^k \sum_{e\in E^i} p(e) \mathbbm{1}[s_{j-1}\leq v(e) < s_j] \times RRank(e, E^i, [s_{j-1}, s_j))
\end{eqnarray*}
which proves that $\mathbb{E}_S[RR_i]$ can be decomposed into the sum of $k$ parts, where the $k$-th part is independent of the first $k-1$ parts. 

\section{Bounds on $C_n(R)$}

We give the proofs for two theorems which provide some useful insights on the convergence rate and sample complexity of the learning objective function in our second method (Equation~(\ref{eq:cr})). Both theorems leverages the Dvoretzky-Kiefer-Wolfowitz (DKW) inequality~\cite{Karimzadehgan:2010:ETI:1871437.1871631} and the following property:

\newtheorem{property}{Property}
\begin{property}
For any real value sequence $x_1, \cdots, x_n$ and $y_1, \cdots, y_n$:
\begin{eqnarray*}
|\sum_{l=1}^m x_l y_l | \leq \sum_{l=1}^m |x_l| \times \max_{l'} |y_{l'}|
\end{eqnarray*}
\label{th:3}
\end{property}

\newtheorem{theorem}{Theorem}
\begin{theorem}
Given query number $n$, suppose the relevant percentages $z^1, \cdots, z^n$ defined in Section~\ref{sec:problem} are independent and identically distributed. Let $F_n$ denote their empirical CDF: 
\begin{eqnarray*}
F_n(r) = \frac{1}{n}\sum_{i=1}^n \mathbbm{1}[z^i < r]
\end{eqnarray*} 
and $F$ denote the true CDF. Suppose $C_n$ is defined by Equation~(\ref{eq:cr}), and $C$ is the true function of $C_n$:
\begin{eqnarray*}
C(R) = \sum_{j=1}^k \Delta r_j\times (F(r_j) - F(r_{j-1}))
\end{eqnarray*}

If the number of ranges is set to $k$, we can prove that given a constant $\epsilon > 0$:
\begin{eqnarray*}
&&\mathbb{P}[\sup_{R} |C_n(R) - C(R)| > \epsilon] \leq 2e^{-2n\epsilon^2/(k-1)^2}
\end{eqnarray*}
\proof Using Property~\ref{th:3}:
\begin{eqnarray*}
|C_n(R) - C(R)| &=& |\sum_{j=0}^k (F_n(r_j) - F(r_j))\times (\Delta r_j - \Delta r_{j+1})|\\
&\leq & \sum_{j=0}^k |F_n(r_j) - F(r_j)|\\
&=& \sum_{j=1}^{k-1}|F_n(r_j) - F(r_j)|
\end{eqnarray*}
where $\Delta r_0 = \Delta r_{k+1} = 0$. If $\sup_R |C_n(R) - C(R) | > \epsilon$, denote $\argmax_R |C_n(R) - C(R) |$ as $R^0 = [r_1^0, \cdots, r_{k-1}^0]$, therefore:
\begin{eqnarray*}
\sum_{j=1}^{k-1} |F_n(r_j^0) - F(r_j^0)| > \epsilon
\end{eqnarray*}  
For at least one these $j$'s we must have $|F_n(r_j^0) - F(r_j^0)| > \frac{\epsilon}{k-1}$ and thus $\sup_{r\in (0,1)} |F_n(r) - F(r) | > \frac{\epsilon}{k-1}$, therefore
\begin{eqnarray}
&&\mathbb{P}[\sup_R |C_n(R) - C(R)| > \epsilon] \nonumber\\
&\leq & \mathbb{P}[\sup_{r\in (0,1)} |F_n(r) - F(r)| > \frac{\epsilon}{k-1}]\nonumber\\
& \stackrel{\text{DKW}}{\leq} & 2e^{-2n\epsilon^2 / (k-1)^2}\label{eq:th1}\qed
\end{eqnarray}
\label{th:1}
\end{theorem}

Theorem~\ref{th:1} describes the convergence rate of $C_n(R)$ for any point in the simplex space $\Delta^k$. As $k$ increases, bound (\ref{eq:th1}) becomes looser. However, under certain setting, this bound will not increase with $k$. We can show it with Theorem~\ref{th:2} and Theorem~\ref{th:3} below:

\begin{theorem}
\label{th:asymptotic}
Suppose we have the same setting as Theorem~\ref{th:1}, but in addition, the true CDF $F$ is strongly concave. Denote $\argmin_R C(R)$ as $R^* = [r_1^*, \cdots, r_{k-1}^*]$, then the widths of $R^*$ is monotonously non-decreasing:
\begin{eqnarray}
\Delta r_1^* \leq \Delta r_2^* \leq \cdots \leq \Delta r_k^*\label{eq:mono}
\end{eqnarray}

\proof Since $F$ is strongly concave, for any $R$ and any pair of adjacent ranges $[r_j, r_{j+1})$ and $[r_{j+1}, r_{j+2})$ in $R$, we have:
\begin{eqnarray}
\frac{\Delta F(r_{j+2})}{\Delta r_{j+2}} < \frac{\Delta F(r_{j+1})}{\Delta r_{j+1}}\label{eq:concave}
\end{eqnarray}
where $\Delta F(r_{j+1}) = F(r_{j+1}) - F(r_{j})$. 

Given the optimal point $R^*$, now consider $R'$, which is same as $R^*$ except for replacing $r_{j+1}^*$ with $(r_j^* + r_{j+2}^*)/2$. Since $R^*$ is the optimal point:
\begin{eqnarray}
C(R^*) \leq C(R')\label{eq:rstar}
\end{eqnarray} 
By canceling the same terms on the L.H.S. and R.H.S. of (\ref{eq:rstar}) we can get:
\begin{eqnarray}
&\Delta r_{j+1}^* \Delta F(r_{j+1}^*) + \Delta r_{j+2}^* \Delta F(r_{j+2}^*)\nonumber\\
&\leq \frac{\Delta r_{j+1}^* + \Delta r_{j+2}^*}{2} (\Delta F(r_{j+2}^*) + \Delta F(r_{j+1}^*))\nonumber\\
\Rightarrow & (\Delta r_{j+1}^* - \Delta r_{j+2}^*) (\Delta F(r_{j+2}^*) - \Delta F(r_{j+1}^*)) \geq 0\label{eq:eq2}
\end{eqnarray}
Suppose (\ref{eq:mono}) is not true, and there exists a $j$ such that $\Delta r_{j+2}^* < \Delta r_{j+1}^*$. It follows from (\ref{eq:concave}) that $\Delta F(r_{j+2}^*) < F(r_{j+1}^*)$, therefore $(\Delta r_{j+1}^* - \Delta r_{j+2}^*) (\Delta F(r_{j+2}^*) - \Delta F(r_{j+1}^*)) < 0$, which contradicts with (\ref{eq:eq2}). \qed
\label{th:2}
\end{theorem}

\begin{theorem}
Suppose we have the same setting as Theorem~\ref{th:1}. In addition, the true CDF $F$ is strongly concave and $R^*=\argmax_R C(R)$. Denote $\mathcal{R}^*$ as a small enough region near $R^*$ where $(\ref{eq:mono})$ stays true, then for constant $\epsilon>0$:
\begin{eqnarray}
&&\mathbb{P}[\sup_{R\in\mathcal{R}^*} |C_n(R) - C(R)| > \epsilon] \leq 2e^{-2n\epsilon^2}\label{eq:concave}
\end{eqnarray}
\proof Following (\ref{eq:mono}), for any $[r_j, r_{j+1})$ in $R\in \mathcal{R}^*$, $|\Delta r_j - \Delta r_{j+1}| =\Delta r_{j+1} - \Delta r_j $. Using Property~\ref{th:3}:

\begin{eqnarray*}
&&\sup_R|C_n(R) - C(R)| \\
&\leq & \sup_R(\Delta r_k - \Delta r_1)\times \max_j |F_n(r_j) - F(r_j)|\\
&\leq & \sup_{r\in (0,1)} |F_n(r) - F(r)|
\end{eqnarray*}

Following DKW inequality we get (\ref{eq:concave}).
\qed
\label{th:3}
\end{theorem}

We may combine the results in Theorem~(\ref{th:1}-\ref{th:3}) with experimental results in Section~\ref{sec:exp} and draw some conclusions. Recall that our second method achieves better experimental results on `Laptop' category than `TV'. From Figure~\ref{fig:laptoptv}, we can observe that the true CDF of `Laptop' is strongly concave while that of `TV' is mostly linear. Meanwhile, Theorem~\ref{th:3} shows that when the true CDF is strongly concave, $C_n(R)$ also has better convergence rates. The two results demonstrate some consistency between theoretical analysis and experimental results. 

\end{document}